\documentclass[twocolumn,showpacs]{revtex4}

\usepackage{graphicx}%
\usepackage{dcolumn}
\usepackage{amsmath}

\newcommand{\be}{\begin{equation}}
\newcommand{\ee}{\end{equation}}
\newcommand{\bey}{\begin{eqnarray}}
\newcommand{\eey}{\end{eqnarray}}
\newcommand{\ba}{\begin{array}}
\newcommand{\ea}{\end{array}}
\newcommand{\bi}{\begin{itemize}}
\newcommand{\ei}{\end{itemize}}
\newcommand{\bem}{\begin{enumerate}}
\newcommand{\eem}{\end{enumerate}}
\newcommand{\bw}{\begin{widetext}}
\newcommand{\ew}{\end{widetext}}

\newcommand{\pp}{\partial}

\newcommand{\ww}{\widetilde}

\newcommand{\bP}{{\bf P}}
\newcommand{\br}{{\bf r}}
\newcommand{\bR}{{\bf R}}
\newcommand{\bx}{{\bf x}}
\newcommand{\by}{{\bf y}}
\newcommand{\bV}{{\bf V}}

\begin{document}

 \title{
 Entanglement and Disentanglement, Probabilistic Interpretation of Statevectors,
 and Transformation between Intrinsic Frames of Reference}

\author{Wen-ge Wang$^{\dag}$}
\affiliation{
 Department of Physics and Centre for Computational Science and Engineering,
 National University of Singapore, 117542 Singapore
 \\ Department of Physics, South-east University, Nanjing 210096, China
 }

 \date{\today}

 \begin{abstract}

 We study a quantum theory based on two assumptions:
 In the intrinsic frame of reference of an isolated, macroscopic system,
 (i) the system has no global motion and is not entangled with any other system,
 (ii) time evolution of statevectors of systems outside the system satisfy Schr\"{o}dinger equation.
 A process of collision-type interaction between a microscopic system and a macroscopic system
 is studied in an auxiliary frame of reference.
 In transforming the statevector of the two systems obtained in the auxiliary frame of reference
 to the intrinsic frame of reference of the macroscopic system,
 the above first assumption requires a discontinuous change of the statevector.
 A probabilistic interpretation is given to the statevector for the discontinuous change.
 For the microscopic system, the density matrix given
 in the theory here is equal to the reduced density matrix given in the usual quantum mechanics.

 \end{abstract}

 \maketitle



 Could an isolated, macroscopic physical system possess an intrinsic frame of reference (FR)
 in which the system has no global motion?
 Here ''isolated'' means negligible interaction with other systems.
 If the center-of-mass position and the total momentum of the system satisfy the uncertainty
 principle, then, the answer is negative.
 The uncertainty principle reflects the fact that, e.g., the disturbance on a system,
 given by a measurement on the position of the system,
 induces uncertainty in the momentum of the system.
 However, for an observer {\it inside} an isolated system, the law of conservation of momentum implies that
 the uncertainty principle is not applicable to the center-of-mass position and the total
 momentum of the system.
 This is because measurement performed inside the system does not change the total momentum
 of the system
 and, as a result, the uncertainty in the center-of-mass position of the system can in principle
 be made smaller and smaller, without disturbing the total momentum.

 One assumption made in this paper is that some type of isolated system may have an intrinsic FR
 in which the system has no global motion.
 No global motion means that the center-of-mass of the system has a definite and fixed position
 in its intrinsic FR, hence, behaves classically.
 The internal motion of the system can be assumed as behaving quantum mechanically.
 The classical feature of the center-of-mass motion denies possible quantum link between the internal
 quantum motion of the system and the quantum motion of an external system described in the intrinsic FR.
 For this reason, we assume further that in its own intrinsic FR the system is not entangled
 with any external system.

 Then, a natural question is whether the above assumption may lead to results in confliction with
 statistical predictions of quantum mechanics that have been confirmed experimentally.
 In this paper, we show that the confliction is avoidable.
 Specifically, we propose a quantum theory with the above assumption as a basic assumption and
 study a process in which a microscopic system and a macroscopic system have collision-type interaction.
 We find that the statistical description for the microscopic system in the theory here
 is the same as that given in the usual quantum mechanics.
 Hence, quantum mechanics modified in this way may keep its predicting ability.

 Furthermore, the basic assumption mentioned above leads to several interesting consequences,
 which are different from the usual quantum mechanics.
 First, if an isolated system is entangled with another system in some FR, when the statevector
 of the two systems in the FR
 is transformed to the intrinsic FR of the isolated system, the entanglement is required to be
 broken and a discontinuous change of the statevector takes place.
 The difference between this discontinuous change of statevector and wavepacket reduction in the
 formalism of standard quantum mechanics (see, e.g., \cite{Neumann-qm,BG03}),
 lies in that it is a result of the transformation between the two FRs, not a dynamical process.
 This supplies a new approach to the measurement problem, which is still a debating
 topic and have received renewed interest in recent years (\cite{Laloe01,Zurek03,Schloss04,BG03}).
 The measurement problem is raised by the relation
 between the two dynamical principles for statevectors in the formalism of standard quantum mechanics,
 namely, Schr\"{o}dinger evolution and wavepacket reduction in measurement processes.
 Second, the statevectors in the two FRs for the two systems have a probabilistic relation.

\vspace{0.2cm}
 \noindent {\it Notations used in the paper} ---

 We use $A,B,...$ to denote macroscopic systems possessing intrinsic FRs and $S$ to denote a microscopic system.
 The intrinsic FR attached to $A$ is denoted by FR$_A$, and so on.
 For $K=A,B,S,...$, the symbol $\bx_K$ (sometimes $\by_K$) is used to indicate the collection of the coordinates
 of all the particles in the system $K$,
 $\bR_K$ is for the center-of-mass coordinates of $K$, and $\br_K$ for the collection of the
 relative coordinates of the particles in $K$.
 Hence, $ \bx_K = (\bR_K, \br_K)$.
 We use $\Psi_K(\bx_K)$  to indicate the wavefunction  of a system $K$,
 $\Phi_K(\bR_K)$  for its center-of-mass coordinates part,
 and $\phi_K(\br_K)$  for its relative coordinates part.
 When two FRs, FR$_A$ and FR$_B$, are employed, descriptions made in FR$_B$ are indicated by
 primes and there is no prime for descriptions made in FR$_A$.

 For simplicity in discussion, we do not consider self-rotational motion, or spin of systems
 and we consider only non-relativistic systems.

\vspace{0.2cm} \noindent {\it Basic assumptions} ---

 We make two basic assumptions:
 \bi \item
 Assumption-I. Some type of isolated, macroscopic system $A$ has an {\it intrinsic FR},
 in which $A$  has no global motion and has an internal motion
 described by a wavefunction $\phi_A(\br_A)$.
 \ei
 No global motion means that $\bR_A$ does not move and $\bP_A=0$ in FR$_A$.
 As a result of this assumption,
 in {\it its own intrinsic FR}, a macroscopic system $A$ is not entangled with any other system.
 Otherwise, the state of $A$ in FR$_A$ is not of the form $\phi_A(\br_A)$.
 This is not in confliction with the phenomenon of entanglement of,
 e.g., pairs of photons, which has been soundly confirmed by experiments (see, e.g.,
 the review paper \cite{Zeilinger99rmp}),
 because the Assumption-I concerns macroscopic systems only.

 The second basic assumption is:
 \bi  \item
 Assumption-II. In FR$_A$ of an isolated, macroscopic system $A$,
 time evolution of the state of any isolated system $K$ outside $A$ satisfies Schr\"{o}dinger equation,
 \be \label{S-eq} i\hbar \frac{\pp }{\pp t} \Psi_K(\bx_K,t) = H_K \Psi_K(\bx_K,t), \ee
 where $H_K$ is the Hamiltonian of system $K$.
 The internal state of $A$, $\phi_A(\br_A)$ satisfies a similar Schr\"{o}dinger equation
 with $H_K$ replaced by the internal Hamiltonian.
 \ei


\vspace{0.2cm}
 \noindent  {\it Collision-type interaction process} ---

 For a system $A$ in interaction with a system $S$, the above assumptions do not tell whether
 $A$ has an intrinsic FR or not.
 To describe the interaction process, we can employ the intrinsic FR of another macroscopic system.

 To simplify the discussion, we consider the following collision-type interaction
 between $S$ and $A$:
 In the initial period of time, the system $S$ has negligible interaction with $A$
 and $A$ has an intrinsic FR;
 in the following interaction period of time, the two systems have non-negligible interaction;
 in the final period of time, $S$ leaves and does not interact with $A$ anymore,
 as a result $A$ has its intrinsic FR.
 The problem we are to study is, if in the initial period of time the states of $S$ and $A$
 in FR$_A$ are known,
 then, what is the prediction for the states of the two systems in FR$_A$ in the final period of time?

 To describe the interaction between $S$ and $A$, we employ an auxiliary FR$_B$ of an
 isolated system $B$.
 In FR$_B$, by the Assumption-II, time evolution of the two interacting systems $S$ and $A$
 satisfies Schr\"{o}dinger equation.

\vspace{0.2cm}
 \noindent  {\it FRs with almost definite relative position and velocity} ---

 Before studying the interaction process, we need to transform the states of $S$ and $A$
 in FR$_A$ to FR$_B$.
 We assume that in FR$_B$ the center-of-mass motion of $A$ is described by $\Phi_A'(\bR_A')$,
 which is a quite narrow wavepacket such that the center-of-mass of $A$ has almost definite
 position and velocity at each time in the whole time period considered.

 Let us first consider a narrow Gaussian wavepacket,
 \bey g(\bR_A')= (\pi \sigma^2)^{-3/4} e^{ i \bR_A'  \cdot  \bP_0' / \hbar }
 \exp \left [ -\frac {(\bR_A' - \bR_0')^2}{2\sigma^2}  \right ], \label{gaussian}  \eey
 where $ \bR_0'$ and $ \bP_0' $ are the centers of the wavepacket in the
 position and momentum spaces, respectively.
 The dispersions of the Gaussian wavepacket in $\bR_A'$ and $\bV_A' \equiv \bP_A' / M_A$
 are $\sigma $ and $\xi = \hbar /  \sigma M_A$, respectively.
 Using $[m]$ to denote the unit of mass and defining a dimensionless quantity $W=M_A/[m]$,
 if the dependence of $\sigma $ and $\xi$  on $M_A$ is set as
 $ \sigma , \ \xi \propto {1}/{\sqrt{W}}$,  in the limit $M_A \to \infty $,
 we have $ \sigma , \ \xi \to 0$.


 Generally, it is unnecessary for $\Phi_A'(\bR_A')$ to be a Gaussian wavepacket.
 As long as the dispersions in $\bR_A'$ and $\bV_A'$ go to zero in the limit $M_A \to \infty$,
 for a system $A$ with sufficiently large mass,
 FR$_A$ can be regarded as having a classical-type relation to FR$_B$.
 Since FR$_B$ is just an auxiliary FR, we can choose $\bP_0'=0$ and $\bR_0'=0$,
 then, there is no relative motion between the two FRs.
 Note that in the usual quantum mechanics, two such FRs can be regarded as identical,
 however, here as a result of the Assumption-I they are not.


 Suppose the states of $S$ and $A$ in FR$_A$ are $\Psi_S(\bx_S)$ and $\phi_A(\br_A)$,
 respectively.
 The state of the big system $S+A$ in FR$_B$ is then
 \bey   \Psi_{S+A}' =  \Phi_A'(\bR_A')  \phi_A(\br_A') \Psi_S(\bx_S'). \label{tran-AB} \eey


\vspace{0.2cm}
 \noindent  {\it Interaction process described in FR$_B$} ---

 In FR$_B$, the total Hamiltonian of the big system $S+A$ has the form
 \be H_T' =  H_A'+H_I'  +H_S', \ \ \text{with} \ H_A' = \frac{(\hat{\bP}_A')^2}{2M_A}
 + H_{A{\rm in}}', \label{H-total-non}  \ee
 where $H_{A{\rm in}}'$ is the internal Hamiltonian of the system $A$,
 $\hat{\bP}_A'= -i\hbar \pp / \pp \bR_A'$, $H_I'$ is the interacting Hamiltonian as a
 function of $(\bx_S'-\bR_A',\br_A')$, and $H_S'$ is the Hamiltonian of $S$.
 Time evolution of the state $\Psi_{S+A}'$ is given by Schr\"{o}dinger equation,
 with an initial state $\ww \Psi_{S+A}'$ of the form on the right hand side of Eq.~(\ref{tran-AB}),
 \be  \Psi_{S+A}'(t) = e^{-iH_T't/\hbar } \ww \Psi_{S+A}'. \label{Psi-1} \ee
 Here we use tilde to indicate initial condition.
 Approximately, $\Psi_{S+A}'(t)$ has the following form
 \bey \Psi_{S+A}'(t) \simeq \Psi_{S+A}^{app'}(t) \equiv \Phi_A'(\bR_A',t) \Psi_1'(t) , \label{Psi-SA-AB-t} \eey
 where
 \bey \Phi_A'(\bR_A',t) = e^{-i{(\hat{\bP}_A')^2t}/{2M_A}\hbar }
 \ww \Phi_A'(\bR_A'), \hspace{0.2cm}
 \\  \Psi_1'(t) = e^{-i(H_I' +H_S'+ H_{A{\rm in}}')t/\hbar }
 \ww \phi_A(\br_A') \ww \Psi_S(\bx_S'). \label{Psi-SA-real} \eey
 To prove this, first note that
 \be i \hbar \frac{\pp }{\pp t} \Psi_{S+A}^{app'}(t) = H_T' \Psi_{S+A}^{app'}(t)
 - \Phi_A'(\bR_A',t) \frac{(\hat{\bP}_A')^2}{2M_A} \Psi_1'(t), \label{app-2}  \ee
 where $(H_I' +H_S'+ H_{A{\rm in}}')$ being commutable with $\bR_A'$ has been used.
 For a micro system $S$, the interaction between $S$ and $A$ has an energy scale negligibly small
 compared with that of $A$.
 Hence, since the dependence of $\Psi_1'(t)$ on $\bR_A'$ is given by the interaction $H_I'$,
 we have
 \be \frac{1}{2M_A} (\hat{\bP}_A')^2 \Psi_1'(t) \sim 0, \label{app-0} \ee
 due to the large denominator $M_A$.
 This means the influence of the interaction between $A$ and $S$
 on the center-of-mass motion of $A$ is negligible.
 Substituting Eq.~(\ref{app-0}) into Eq.~(\ref{app-2}), we prove Eq.~(\ref{Psi-SA-AB-t}).


\vspace{0.2cm}
 \noindent  {\it States of $S$ and $A$ in FR$_A$ in the final period of time} ---

 To transform, in the final period of time, from FR$_B$ to FR$_A$, we first write
 the explicit dependence of $\Psi_1'(t)$ in Eq.~(\ref{Psi-SA-real}) on its variables, which is
 $ \Psi_1' (\bx_S'-\bR_A',\bx_S',\br_A',t)$, where the dependence on $(\bx_S'-\bR_A')$
 is given by $H_I'$.
 Due to the narrowness of $\Phi_A'(\bR_A')$, $\bR_A'$ in $\Psi_1'$ can be approximated by
 $\bR_0'=0$, hence, we have
 \be \Psi_{S+A}'(t) \simeq \Phi_A'(\bR_A',t) \Psi_{1}' (\bx_S',\br_A',t). \label{Psi-SA-app}  \ee
 Clearly $\Phi_A'(\bR_A',t)$ does not appear in the description in FR$_A$.
 For $\Psi_{1}' (\bx_S',\br_A',t)$, we first transform it to
 \be \Psi_{1} (\bx_S,\br_A,t) \equiv \Psi_{1}'(\bx_S,\br_A,t).  \ee
 The statevector $\Psi_{1} (\bx_S,\br_A,t)$ can not be interpreted as the state of $S+A$ in FR$_A$,
 because it is generally an entangled state, while the Assumption-I requires
 a product state for  $S$ and $A$ in FR$_A$.

 To avoid confliction with the Assumption-I, it seems the only way is to
 give $\Psi_{1} (\bx_S,\br_A,t)$ a probabilistic interpretation, by making use of its expansion
 in product states.
 For this, the most natural method is to use the Schmidt decomposition \cite{nc-book}
 of $\Psi_{1} (\bx_S,\br_A,t)$ in orthogonal, normalized states,
 \be \Psi_{1} (\bx_S,\br_A,t) =
 \sum_j C_j \Psi_{Sj}(\bx_S,t) \phi_{Aj}(\br_A,t). \label{eq-j-2} \ee
 The interpretation of this equation is that in FR$_A$ there is a probability $|C_j|^2$
 for the state of $A$ to be $\phi_{Aj}(\br_A,t)$ and the state of $S$ to be $\Psi_{Sj}(\bx_S,t)$.
 Hence, $S$ is in a mixed state in FR$_A$.
 Entanglement is then not an absolute property of systems, but depends on the FR taken.

 Hence, in going from FR$_B$ to FR$_A$, the statevector has a discontinuous change in the transformation,
 namely, from $\Psi_{1}$ to one of the possible $\Psi_{Sj} \phi_{Aj}$.
 This is different from wavepacket reduction in the formalism of standard quantum mechanics,
 which is a dynamical process.
 Note that this transformation is usually irreversible.
 The discontinuous change of statevector implies that information in different intrinsic FRs can
 be nonidentical.
 This is not as strange as at first sight, if one notes that the system $A+S$ and the system $B$
 are isolated in the time period considered.
 Indeed, there is no experimental evidence for that information
 in two FRs attached to two isolated systems must be identical.

 The possibility for two $|C_j|$ in Eq.~(\ref{eq-j-2}) to be identical is quite rare.
 If it happens, Schmidt decomposition for the related components is not unique.
 This problem may be solved by a natural extension of a similar case with
 small difference in two $|C_j|$; if a natural extension does not exist,
 further probabilistic interpretation for possible Schmidt decomposition is needed.


 {\it Density matrix} --- In the final period of time, $S$ is in a mixed state in FR$_A$,
 described by the density matrix
 $\rho_S(\by_S,\bx_S,t) = \sum_j |C_j|^2 \Psi_{Sj}^*(\by_S,t)  \Psi_{Sj}(\bx_S,t)$.
 In the usual quantum mechanics, one has the same time evolution of $\Psi_{S+A}'(t)$ as in
 Eq.~(\ref{Psi-1}), hence, for $S$ one has the reduced density matrix
 \be \rho^{re}_S(\by_S',\bx_S',t) = \int d\bx_A' {\Psi_{S+A}'}^*(\by_S',\bx_A',t)
 \Psi_{S+A}'(\bx_S',\bx_A',t) . \ee
 Making use of Eqs.~(\ref{Psi-SA-app}-\ref{eq-j-2}), taking  the limit $M_A \to \infty $,
 and noting that FR$_A$ and FR$_B$ are identical in the usual quantum mechanics,
 it is easy to verify that $\rho_S = \rho^{re}_S$,
 hence, the theory here and the usual quantum mechanics predict the same density matrix
 for the state of $S$.

 Extension of the above results is straightforward for the case in which $S$ is
 partly absorbed by $A$.
 Suppose in the final period of time $S$ is divided into two parts $S_1$ and $S_2$,
 with $S_1$ absorbed by $A$ forming a combined system $S_1+A$, denoted by $A_1$,
 and $S_2$ far from $A$.
 In the state $\Psi_{1}$, this means that $|\Psi_{1}|^2$ is negligibly small
 if either $S_1$ is far from $A$ or $S_2$ is close to $A$.
 Then, for the two systems $S_2$ and $A_1$, one can make a decomposition like Eq.~(\ref{eq-j-2})
 and give similar interpretation in the intrinsic FR of $A_1$.
 More generally, combining the two cases, we have
 \bey  \Psi_{1} = \sum_j C_j \Psi_{Sj} \phi_{Aj}
  + \sum_k D_k \Psi_{S_2,k} \phi_{A_1,k}, \label{eq-j-3} \eey
 with similar interpretation.

\vspace{0.2cm}
 \noindent {\it An application of the theory: a position measurement} ---

 Consider a special case of $S$ being partly absorbed by $A$,
 in which $S$ is composed of two noninteracting particles $a$ and $b$ in the initial period of time.
 The particle $a$ interacts with the system $A$ in the interaction period of time
 and is finally absorbed by $A$, forming a combined system $a+A$, the effect of which may be observed;
 while the particle $b$ does not interact with the system $A$ or the particle $a$
 in the whole period of time considered.

 We assume that initially $S$ is in the following entangled state
 as a result of previous interaction, in Schmidt decomposition in normalized states,
 \be \ww \Psi(\bx_S) = \sum_{l=1}^2 c_l \ww \Psi_{a,l}(\bx_a) \ww \Psi_{b,l}(\bx_b), \label{is-ef} \ee
 where the two components $\ww \Psi_{a,1}(\bx_a)$ and $\ww \Psi_{a,2}(\bx_a)$ are well separated in space
 and remain well-separated in the whole initial period of time.
 The interaction process is studied in the auxiliary FR$_B$.
 Substituting Eq.~(\ref{is-ef}) into Eq.~(\ref{tran-AB}), then, making use of
 Eqs.~(\ref{Psi-1}-\ref{Psi-SA-real}), we find that $\Psi_1'$ in Eq.~(\ref{Psi-SA-app}) has the following form
 \be \Psi_1'(t) = \sum_l c_l \phi_{A+a,l}'(\br_{A+a,l}',t)  \Psi_{b,l}(\bx_b',t), \label{ef} \ee
 where the components of particle $b$ evolve under its own Hamiltonian
 and $\phi_{A+a,l}'$ is the time evolution of $\ww \phi_A \ww \Psi_{a,l}$.
 For this process to be able to serve as a position measurement, we assume that
 in the two components $\phi_{A+a,l}'$ of $l=1,2$ the particle $a$ is localized in different spatial regions.
 Then, $\phi_{A+a,l}'$ of $l=1,2$ are orthogonal
 and the right hand side of Eq.~(\ref{ef}) is a Schmidt decomposition, at least approximately.
 Consequently, in FR$_A$, the combined system $A+a$ is in a mixed state, either in $\phi_{A+a,1}'$
 or in $\phi_{A+a,2}'$, with the probability $|c_1|^2$ and  $|c_2|^2$, respectively.

 There is another important consequence of the above interaction process. That is,
 in the final period of time the particle $b$ is also in a mixed state and the entanglement of the
 two particles $a$ and $b$ is broken in FR$_A$.

 Extension of the above results to the case of more than two
 far-separating components of the particle $a$ in Eq.~(\ref{is-ef})
 and to the case in which $a$ and $b$ are composed of more than one particles are straightforward.
 It is seen that the initial entanglement of $a$ and $b$ plays an important role in the above
 explanation of position measurement.
 We remark that initial entanglement is a general phenomenon.
 In fact, to prepare $a$ in an initial state, it is necessary to have $a$ interact with some other
 system, which inevitably induces initial entanglement.

\vspace{0.2cm}
 \noindent {\it Discussions} ---

 Decoherence induced by environment has been extensively studied in recent
 decades \cite{Zurek03,Zeh70,Zeh05,Zurek81,PZ99,Zurek03-prl,BHS01,Eisert04},
 revealing the emergence of classical properties
 in open quantum systems due to unavoidable coupling to their environment.
 The decoherence programm can not completely solve the quantum measurement problem yet \cite{Schloss04}.
 In particular, to solve the problem of definite outcomes in quantum measurement,
 some type of interpretation of quantum mechanics is needed,
 e.g., the many-worlds interpretation proposed by Everett \cite{Everett57}.
 An advantage of the theory here is that its basic Assumption-I supplies a physical mechanism to
 solve the problem of definite outcomes,
 as illustrated in the above discussion for a simple position measurement scheme.

 Since the two systems $S$ and $A$ discussed above have no interaction
 in the final period of time, it is simple to see that the states of $S$ and $A$ in FR$_A$ in the final
 period of time are in agreement with pointer states in the theory of decoherence \cite{Zurek81,Zurek03}.
 In a further development of the theory here,
 it seems plausible to expect that the Assumption-I could be generalized to the case of $A$ having weak
 interaction with $S$.
 In this case the macroscopic system $A$ can be regarded as an environment of the microscopic system $S$.
 In FR$_B$, the decoherence theory predicts pointer states as a result of the interaction.
 This should be useful in the development of the theory here.
 Other directions of development of the theory may include more general interaction situations,
 as well as other types of FR, e.g., FRs attached to mesoscopic even microscopic systems.

 In the usual quantum mechanics, the concept of FR is used almost in the same way as in
 classical mechanics.
 Presently, little is known for the influence of quantum motion of systems on properties of
 the attached FRs.
 The theory discussed above shows that intrinsic FRs may have unexpected properties,
 when the quantum motion of the attached systems are considered.

   This work was supported in part by the Natural Science Foundation of China Grant No.10275011.

 \end{document}